\documentstyle[prd,preprint,aps]{revtex}
\begin{document}

\draft
\preprint{JFM-PUB-001}

\title{An Isobar Model of ${e^+ e^- \rightarrow \pi^+ \pi^- %
\pi^0 \pi^0}$}
\author{John F. McGowan, Ph.D.\\University of Illinois
at Urbana-Champaign}
\address{1379 Snow Street \#3, Mountain View, CA, 94041 }
\date{\today}
\maketitle

\begin{abstract}
This paper presents several matrix elements
for processes that may contribute to \({e^+
e^- \rightarrow \pi^+ \pi^- \pi^0 \pi^0}\)
derived using the isobar model and
the Lorentz invariant amplitude method.  The formulas
may be used to measure the electromagnetic
form factors of the \(\rho\) meson in experimental data.
Quantum Chromodynamics ({\bf QCD}) calculations predict
the electromagnetic form factors
of the \(\rho\) meson which can be compared to the measured values.
The formulas in this paper
may be used to study the \(a_1\) meson
if \({ e^+ e^- \rightarrow a_1 \pi }\)  contributes to \({e^+
e^- \rightarrow \pi^+ \pi^- \pi^0 \pi^0}\).
\end{abstract}

\pacs{23.23.+x, 56.65.Dy}

\tableofcontents

\section{Introduction}

The isobar model (described below) is a phenomenological
theory of intermediate energy (a few $GeV$) strong interactions
\cite{ft:Watson-1,ft:Watson-2,ft:Cook,ft:Ball,ft:Mandelstam,%
ft:Fleming,%
ft:Freedman,ft:Morgan}.  Phenomenology is
``a descriptive or classificatory account of the phenomena of a given
body of knowledge, without any further attempt at explanation''
\cite{ft:Webster}.
In this sense, the isobar model is phenomenology.
Quantum Chromodynamics ({\bf QCD}) hopefully
provides a fundamental explanation of the interactions described
by the isobar model.
The isobar model cannot be derived at present
from {\bf QCD}. It presumably is
not a fundamental explanation of strong interactions, but
provides a description of many phenomena in strong interactions.

This paper derives formulas for the isobar model
processes that may contribute to the reaction ${e^+ e^- \rightarrow
\pi^+
\pi^- \pi^0 \pi^0}$.  These formulas can be compared to ${e^+ e^-
\rightarrow \pi^+ \pi^- \pi^0 \pi^0}$ data from experiments such as
CLEO, CLEO II, the Beijing Electron Spectrometer (BES), {\bf Mark III
}, and {\bf DM2}.

The formulas can be used to measure
the electromagnetic form factors of the $\rho$ meson in
experimental data on ${e^+ e^-%
\rightarrow \pi^+ \pi^- \pi^0 \pi^0}$.  There are at least
two predictions of the electromagnetic form factors of the \(\rho\)
based on {\bf QCD} arguments to compare to the experimental
measurements
\cite{ft:LePage,ft:Brodsky,ft:Chernyak-1,ft:Chernyak-2}.  The isobar
model formulas can be used to determine the properties of the $a_1$
resonance if $e^+ e^-
\rightarrow a_1 \pi$ is a significant contributing process to
${e^+ e^- \rightarrow \pi^+ \pi^- \pi^0 \pi^0}$.  A more detailed
discussion of these results can be found in the thesis on which this
paper is based\cite{ft:McGowan}.

\section{The Isobar Model}

The isobar model assumes that particle production
and decay proceeds via
resonances.  For example, in the process $ \gamma p \rightarrow \pi^0
p $, pion photoproduction, the isobar model assumes the matrix
element is of the general form

\begin{equation}
\label{eq:IsobarModel}
M = \sum_i A_i ME_i(\epsilon_{\gamma}, \epsilon_{p})
{ 1 \over { m^2 - m_i^2 -i \Gamma_i m }}
\end{equation}

where $m_i$ is the mass of resonance {\it i} and $ \Gamma_i $ is the
width of the resonance.  This is the Breit-Wigner formula which
originated in studies of atomic and nuclear physics.
The isobar model uses several variants of the Breit-Wigner formula.
The Breit-Wigner formulas are the non-relativistic
Breit-Wigner formula with constant mass $m_0$ and width $\Gamma_0$,
the relativistic Breit-Wigner formula with constant mass and width,
and Breit-Wigner formulas where the width $\Gamma(m)$ is
a function of the mass.

If the resonance is narrow, there is little difference
between a Breit-Wigner formula with a constant width $\Gamma_0 =
\Gamma(m_0)$ and a Breit-Wigner formula with a width that is a
function of the mass.  If the resonance is broad, the Breit-Wigner
formula with mass-dependent
width $\Gamma(m)$ may be significantly different from a
Breit-Wigner with constant
width.  Additional assumptions are required to specify the
mass-dependent width $\Gamma(m)$.

\(ME_i(\epsilon_{\gamma}, \epsilon_p) \) is a Lorentz invariant
matrix element incorporating the spin and momentum dependence
of the interaction.
$A_i$ in Equation~\ref{eq:IsobarModel} could be a
complex function of the mass $ m^2 = (p_{\gamma} + p_p)^2 $ but it is
generally a complex constant.  This is certainly a
reasonable approximation if the resonance $i$ is narrow.  If the
resonance is broad, this approximation may be invalid.

Pion photoproduction is presumed to proceed by
$$ \gamma p \rightarrow
\sum_I {\rm Resonance \> I } \rightarrow \pi^0 p $$

In the isobar model, the S matrix \cite{ft:Wheeler,ft:Heisenberg} is
the sum of products of Breit-Wigner factors, each
corresponding to a resonance,
and complex matrix elements incorporating spin-dependence.
Physically, this means the interaction
proceeds by sequential two-body interactions.  For example, $ A + B +
C \rightarrow D + E + F $ might proceed by
$ A + B \rightarrow D + X $
followed by $ C + X \rightarrow E + F $.  $X$ is a resonance and
appears as a term $$ 1 \over { (p_A + p_B - p_C)^2 - M_X^2 + i
\Gamma_X M_X } $$ The process might proceed by
$ A + C \rightarrow D +
Y $ and $ B + Y \rightarrow E + F $ as well, where \(Y\) is a
different resonance.

For particle decay, the isobar model is equivalent to
{\it sequential two-body decay } of resonances.  For example, the
decay $ A \rightarrow
B + C + D $ is actually $ A \rightarrow X + D $
followed by
$ X \rightarrow B + C $.  The intermediary resonance $X$ appears
as a Breit-Wigner term with mass $m_X$ and width $\Gamma_X$ in the
S matrix.

The Breit-Wigner formula is
an assumption of the
isobar model.  This formula originated of atomic and
nuclear physics.  Narrow peaks in cross-sections occur in atomic and
nuclear physics.  These are typically interpreted
as bound states in a
central potential.  The Breit-Wigner formula was derived in this
context by applying the mathematics of a
forced damped oscillator such as a tuning fork to
atomic and nuclear systems.

When resonances were discovered in pion-nucleon
scattering, an analogy
was made to the much narrower states observed in atomic and nuclear
physics.  As part of this analogy, the Breit-Wigner formula was
applied to low-energy hadronic interactions.

The isobar model incorporates a Lorentz invariant matrix element
to represent the angular distributions in the interaction due to
the spin of the interacting particles.  The evaluation of these
matrix elements comprises the bulk of this paper.

The isobar model assumes that the resonances have the
same properties --- the same mass, the same width, the
same charge, the
same quantum numbers --- in all processes.  The mass and width refer
to the parameters of the Breit-Wigner formula for the resonance.  The
isobar model assumes a fundamental
spectrum of resonances with
unique properties that describe all low-energy hadronic interactions.
The same resonances occur in $e^+ e^-$ annihilation, pion-nucleon
scattering, and other processes.  In practice, this
spectrum is finite
because only enough energy is present to create the lower energy
resonances.

Generally, isobar models assume that only resonances below a certain
cut-off angular momentum contribute to low-energy nuclear processes.
Resonances with a spin above $2$ are usually
ignored in isobar models.
The higher spin resonances presumably have larger masses.
This makes
intuitive sense if the resonances are excited states of quarks in a
potential.  The energy to create the high spin resonances is not
available.  In practice, radial excitations are frequently
ignored for the same reason.

In summary, the isobar model assumes:

\begin{itemize}

\item The S matrix should be a Lorentz invariant function of initial
and final state 4-momenta and spins.

\item The S matrix can be represented as the sum of products of
Breit-Wigners which represent known resonances.
Physically, this can be
thought of as representing the process as a sequence
of two-body interactions.

\item Resonances are represented by the Breit-Wigner formula.
Frequently, the relativistic Breit-Wigner formula with constant mass
and width is used.

\item The resonances are the same --- have the same mass,
width, charge, and
other quantum numbers --- in all processes: $e^+ e^-$ annihilation,
pion-nucleon scattering, kaon-nucleon scattering, pion
photoproduction,
pion electroproduction, and so forth.

\item Each two-body process has an associated Lorentz invariant
matrix element that represents the angular distributions in the
two-body process due to the spins of the interacting particles.
The rules for constructing these matrix elements are described
below.  The helicity formalism is an alternate scheme for
constructing these matrix elements that is not used in this
paper.

\item Each two-body process has an associated
amplitude which is treated as a complex constant.  For spinless
particles, the most general Lorentz invariant form of this amplitude
for a two-body process would be an arbitrary function of the two-body
invariant mass in the two-body process.  A complex constant is a
reasonable approximation for narrow resonances.  For broad
resonances,
it constitutes an {\it assumption}.

\end{itemize}

In practice, only resonances with less than a cut-off spin \(J\)
are used and radial excitations are frequently ignored.

\section{Lorentz Invariant Amplitude Method}

Lorentz invariant matrix elements are constructed by describing a
decay as sequential two-body or three-body decay vertices.  In
${e^+ e^- \rightarrow \pi^+ \pi^- \pi^0 \pi^0}$,
 $\rho \rightarrow \pi
\pi$ is a two-body vertex and $\omega \rightarrow \pi \pi \pi$ is a
three-body vertex.  The
total matrix element is the product of the matrix elements for each
vertex.  These matrix elements are constructed
from the 4-momenta and
polarization 4-vectors of the in-going and out-going states.

In this paper, the following notation will be used.  A
4-vector is represented as follows:

\begin{equation}
 a_{\mu} =  a = (a_x,a_y,a_z,a_t)
\end{equation}

and

\begin{equation}
 a^{\mu} = (-a_x,-a_y,-a_z,a_t)
\end{equation}

A 4-momentum is:

\begin{equation}
 p_{\mu} =  p = (p_x,p_y,p_z,E)
\end{equation}

where $E$ is the energy.

This analysis uses two kinds of 4-vectors.
4-momenta change sign under the parity transformation.  Polarization
4-vectors, which are usually represented by $\epsilon$, do not
change sign under the parity transformation.  The amplitudes
formed by the Lorentz invariant amplitude
formalism must have definite parity properties, determined
by the intrinsic parities of the particles as described below.

The Minkowski metric $g_{\mu \nu}$ is defined as:

\begin{equation}
 g_{\mu \nu} = \left( \matrix{-1 &0 &0 &0 \cr
				 0 &-1&0 &0 \cr
				 0 &0 &-1&0 \cr
				 0 &0 &0 &1 \cr} \right)
\end{equation}

Then, the dot product of two 4-vectors can be represented as:

\begin{equation}
 a_{\mu} b^{\mu} = g_{\mu \nu} a^{\mu} b^{\nu}
\end{equation}

\begin{equation}
 a_{\mu} b^{\mu} = \delta^{\mu}_{\nu} a_{\mu} b^{\nu}
\end{equation}

\begin{equation}
 a_{\mu} b^{\mu} = ( a \cdot  b)
\end{equation}

\begin{equation}
 ( a \cdot  b) = a_t b_t - a_x b_x - a_y b_y
				       - a_z b_z
\end{equation}

\( \delta^{\mu}_{\nu} \) is the Kronecker delta.

Notice, in this notation:

\begin{equation}
  p \cdot  p = M^2
\end{equation}

With the other possible choice of metric,

\begin{equation}
  p \cdot  p = -M^2
\end{equation}

The polarization 4-vectors $\epsilon(\lambda)$ take the
following form in this notation:

For a massive spin one (1) particle with helicity
$\lambda$ with momentum $\bar p$ along the $x$ axis,

\begin{equation}
\label{eq:polar}
\epsilon(\lambda = \pm 1) = \mp (0,1,\pm i,0) / \sqrt{2}
\end{equation}

\begin{equation}
\label{eq:epsilon}
\epsilon(\lambda = 0) = (p,0,0,E)/M
\end{equation}

where M is mass of particle.
or
\begin{equation}
\label{eq:epsilonExact}
\epsilon(\lambda = 0) = (p,0,0,E)/|p|
\end{equation}

For narrow resonances, Equation~\ref{eq:epsilon} and
Equation~\ref{eq:epsilonExact} are equivalent.  However for
Equation~\ref{eq:epsilonExact}, note that if the massive vector is
broad, then $  p \cdot  p \ne M^2 $ in general.
Equation~\ref{eq:epsilonExact} is the correct formula to use in
general.  Equation~\ref{eq:epsilon} is an approximation.  These
polarization 4-vectors are orthonormal.  They satisfy:

\begin{equation}
 \epsilon(\lambda_i) \epsilon(\lambda_j)^* = \delta_{ij}
\end{equation}

Notice that one dots with complex conjugate to get this.

For a massless vector particle (such as a photon):

\begin{equation}
 \epsilon \cdot p = \epsilon_{\mu} p^{\mu} = 0
\end{equation}

The Lorentz invariant amplitudes are also built using the Levi-Civita
tensor ${\epsilon_{\mu \nu \rho \sigma }} $.  The Levi-Civita tensor
is defined by $$ { \epsilon_{\alpha \beta \gamma \delta}} = \cases{
-1,&if $\alpha \beta \gamma \delta$ is an even permutation of
$xyzt$;\cr +1,&if $\alpha \beta \gamma \delta$ is an odd permutation
of $xyzt$; \cr 0,&otherwise \cr} $$ and $$ \epsilon^{\alpha \beta
\gamma \delta} = - { \epsilon_{\alpha \beta \gamma \delta}} $$

The Levi-Civita tensor changes sign under the parity transformation.
It has some other properties which will be discussed in the section
below on the $\omega \pi$ model.

The Levi-Civita tensor enters into the amplitudes in forms

$$ {\epsilon_{\mu \nu \rho \sigma }}
A^{\mu} B^{\nu} C^{\rho} D^{\sigma} $$

The Levi-Civita tensor is a determinant or a signed
volume of a box defined by four 4-vectors.
If a 4-vector is repeated in the above form (e.g. $A=B$) then
the form is $0$.

The matrix elements are built out of the 4-momenta, polarization
4-vectors, Breit-Wigners, and the Levi-Civita tensor $\epsilon_{
\mu \nu \rho \sigma} $.  The polarization of the intermediate states
must be summed over using the completeness relation.

\begin{equation}
 \sum_{\lambda} \epsilon_{\mu}^{\lambda *}
		  \epsilon_{\nu}^{\lambda} =
		  -g_{\mu \nu}
		  + {p_{\mu} p_{\nu} \over p^2}
\end{equation}

where $\epsilon$ is the polarization 4-vector of the intermediate
state, $p_{\mu}$ is the 4-momentum of the intermediate state,
and $M$ is the pole-mass of the intermediate state.  $g_{\mu \nu}$
is the Minkowski tensor.  For a narrow resonance, $p^2$ can be
replaced by $M^2$ in the completeness relation.  This analysis
involves the $\rho$ which is broad.

In practice, the completeness relation appears in forms of the
type

\begin{equation}
\label{eq:spinsum}
\sum_{\lambda} ( A \cdot \epsilon(\lambda))
		  ( B \cdot \epsilon(\lambda)) =
 { \Bigl[ -( { A} \cdot { B} ) + { { ({ A} \cdot P_{s})
( P_{s} \cdot { B}) } \over { P^2_{s} } } \Bigr] }
\end{equation}

where $ A$ and $ B$ are 4-vectors, either four
momenta or polarization 4-vectors.  $\epsilon(\lambda)$ is the
polarization of an intermediate state which is summed over.
\( P_{s} \) is the momentum of the intermediate state \(s\).

The completeness relation also appears in forms of the type

\begin{eqnarray}
\label{eq:superspinsum}
\sum_{\lambda_a \lambda_b} ( A \cdot \epsilon_a(\lambda_a))
	  (\epsilon_a(\lambda_a) \cdot \epsilon_b(\lambda_b) )
	  (\epsilon_b(\lambda_b) \cdot  B) =
& - \Bigl[ - ( { A} \cdot { B} ) + { {( { A} \cdot P_{b} )
( P_{b} \cdot { B} ) } \over { P^2_{b}} } \Bigr] \nonumber \\
& + { { P_{b} \cdot { B}} \over { P^2_{b}} }{ \Bigl[ -( { A}
\cdot { B}) + { {({ A} \cdot P_{a})(P_{a} \cdot { B}) }
\over {P^2_{a} } } \Bigr]}
\end{eqnarray}

The Lorentz invariant amplitudes usually reduce to formula
constructed of 4-vector products
using the expansions in Equations~\ref{eq:spinsum} and
\ref{eq:superspinsum} above.  This allows one to drop the
Greek indices and provides a
form for the amplitudes which is easy to compute with a 4-vector
dot product function when implementing the model on
a computer.

The specific rules for forming amplitudes (or matrix elements) are:

\begin{itemize}
\item The amplitudes must be formed by contracting
combinations of the available 4-momenta and
polarization 4-vectors.  The
amplitudes must be Lorentz invariant.  4-vectors are not Lorentz
invariant, but the contractions are Lorentz invariants.
All of the polarization 4-vectors {\it must } be used to form the
amplitudes.  The 4-momenta may or may not be used.

\item Different amplitudes for the same process must be linearly
independent to within factors which are rotationally
invariant.  These factors are envisioned as slowly varying
functions of invariant masses.

\item The amplitudes must be either scalars or
pseudoscalars depending
on the required parity properties.  The product of the
intrinsic parities of the decaying particle
and the decay products determines
the required parity transformation properties of the amplitude.

\item Photons must satisfy $ \epsilon^{\mu} p_{\mu} = 0$. $\epsilon$
is the polarization 4-vector of the photon and $p$ is the 4-momentum
of the photon.

\item Intermediate states must be summed over helicities using
the completeness relation above.

\item Intermediate states should include a mass dependent phase.
This is the Breit-Wigner form.

\item If the process being modeled is an eigenstate of $C(G)$ then
the amplitude must be appropriately symmetrized so that the final
state it describes is also an eigenstate of $C(G)$.
$C$ is charge conjugation which transforms a particle into its
anti-particle.  $G$ refers to $G$ parity.

\end{itemize}

\subsection{\bf Mass Dependent Widths }

The Particle Data Book and other sources only give a single
width at the pole mass $\Gamma_0 = \Gamma(m_0) $.  As mentioned
above, the Breit-Wigner form contains a mass dependent width.
For a particle $d$ which decays to $N$ particles, the width is
determined by

\begin{equation}
 \Gamma(d \rightarrow 1, \ldots, n) =
       { 1 \over 2m } \int dLIPS(m^2,P_1,\ldots,P_n) |ME|^2
\end{equation}

where $ME$ is the matrix element of the decay.  $dLIPS$ refers
to integration of Lorentz Invariant Phase Space.

This paper is concerned with two decays: $\rho \rightarrow
\pi \pi$ and $\omega \rightarrow \pi \pi \pi$.  The
volume of phase space changes as the invariant mass of the decaying
particle $d$ changes.  Thus, for this reason alone, the width
is mass dependent.
The isobar model needs formulas that relate $\Gamma(m)$ to
the width $\Gamma_0$ and the mass $m$ of the intermediate state.

For massive spin one (vector) particles decaying to two particles,
the formula

\begin{equation}
 \Gamma(m) = \Gamma_0 { m_0 \over m} ({ p_{cm} \over p_{cm0} })^3
{  { 1 + (R p_{cm0})^2 } \over { 1 + (R p_{cm})^2 } }
\end{equation}

can be used.  It is used for the $\rho$ resonances.  However, it is
inappropriate for the $\omega$ which decays to three pions.  In this
instance, the $\Gamma(m)$ can be
calculated by explicitly integrating the
matrix element squared using the formula for the mass dependent width
above.

In this, $ \sqrt{{{ { 1 + (R p_{cm0})^2 } \over { 1 + (R p_{cm})^2 }
}}} $ are the Blatt-Weisskopf barrier penetration factors
\cite{ft:Blatt}.  The Blatt-Weisskopf barrier penetration factors
represent mathematically that the particles are not point particles.
$R$ is an estimate of the radius of a strongly interacting particle.
Notice that if $R = 0$ then the penetration factors are identically
$1$.  The width $\Gamma$ is the integral of the square of the matrix
element of the decay.  The Blatt-Weisskopf factor $ \sqrt{{{ { 1 + (R
p_{cm0})^2 } \over { 1 + (R p_{cm})^2 } }}} $ appears in the matrix
element.  This relationship between the width and the matrix element
is why the square of the Blatt-Weisskopf barrier penetration factor
appears in the formula for the width.  $p_{cm}$ is the momentum
of the
two-body decay in the resonance rest frame.  The value of $R$ is not
definite.  The value 1 fermi or 5 $GeV^{-1}$ is usually used.

\section{Lorentz Invariant Amplitude Models for ${e^+ e^- \rightarrow
 \pi^+ \pi^- \pi^0 \pi^0}$}

This section contains the main results of this paper, Lorentz
Invariant Amplitudes for several processes that may contribute to
${e^+ e^- \rightarrow \pi^+ \pi^- \pi^0 \pi^0}$.
This paper envisions the
process ${e^+ e^- \rightarrow \pi^+ \pi^- \pi^0
\pi^0}$ as the superposition of several modes listed in
Table~\ref{tab:modes}.  All of the modes except for the last are
represented mathematically as Lorentz invariant complex amplitudes
that interfere.
%
%

The last mode in Table~\ref{tab:modes} is non-resonant four
pions, usually
production of ${ \pi^+ \pi^- \pi^0 \pi^0}$ according to a four-body
phase space distribution added incoherently to the other modes.
Technically, the isobar model precludes this last mode.  It is
incorporated in isobar model analyses as a substitute
for unmodelled modes, detector simulation errors, and so forth
when performing fits to experimental data.

\subsection{\bf The $e^+ e^- \rightarrow \rho^+ \rho^-$ Mode }

This is $V \rightarrow VV $ process where $V$ indicates a vector
particle.  There are three polarization 4-vectors
$\epsilon_{\gamma^*}$,
$\epsilon_{\rho^+}$, and $\epsilon_{\rho^-}$.
\( \epsilon_{\gamma^*}\) is the polarization of
the virtual photon formed
when the electron and positron annihilate.  There are two independent
4-momenta $P$ and $Q$.  $P = (0,0,0,E_{cm}) = P_{\rho^+} + P_{\rho^-}
$ and $ Q = P_{\rho^+} - P_{\rho^-} $.  There are two neutral pions
$\pi^0$.  These are refered to in this analysis as $\pi^0_1$ for the
more energetic pion and $\pi^0_2$ for the less energetic pion.  There
are two ways to form the 4-momenta of the two $\rho$ mesons: $
P_{\rho^+} = P_{\pi^+} + P_{\pi^0_1} $ and $ P_{\rho^-} = P_{\pi^-} +
P_{\pi^0_2} $ or $ P_{\rho^+} = P_{\pi^+} + P_{\pi^0_2} $ and
$ P_{\rho^-} = P_{\pi^-} + P_{\pi^0_1} $.
It will be necessary to add the matrix
elements for these two cases to insure that the final amplitude has
the correct Bose symmetry.
As a first step, the decay of the $\rho$'s
is ignored.

There are three independent amplitudes that can be constructed
according to the rules of the Lorentz invariant formalism.
These are:

\begin{equation}
 ME_1 = (\epsilon_{\gamma^*} \cdot Q) (\epsilon_{\rho^+} \cdot
	   \epsilon_{\rho^-} )
\end{equation}

\begin{equation}
 ME_2 = (\epsilon_{\gamma^*} \cdot Q) (\epsilon_{\rho^+} \cdot
	   Q) (\epsilon_{\rho^-} \cdot Q)
\end{equation}

\begin{equation}
 ME_3 = (\epsilon_{\gamma^*} \cdot \epsilon_{\rho^+})
	  (\epsilon_{\rho^-} \cdot Q) + (\epsilon_{\gamma^*}
	  \cdot \epsilon_{\rho^-}) (\epsilon_{\rho^+} \cdot Q)
\end{equation}

Note that these are the matrix elements at the $ \gamma^*
\rightarrow \rho^+ \rho^- $ vertex only.  The full formulas (see
below) must include the $\rho \rightarrow \pi \pi $ decay.
All of these formulas are parity $-$ and charge conjugation $-$
eigenstates.
The ( $\rho^+
\rho^- $) final state is a charge conjugation
eigenstate \cite{ft:Kramer}.  Kramer
and Walsh's paper quotes these formulas using a different
notation.  Their paper does not consider the $\rho$ decay.

Thus, the total matrix element would be

\begin{equation}
 ME(e^+ e^- \rightarrow \rho^+ \rho^-) =
     A_1 ME_1 + A_2 ME_2 + A_3 ME_3
\end{equation}

where $A_i$ are complex parameters.  The
parameters correspond to the electric,
magnetic, and electric quadrupole form factors of the $\rho$.  The
relations between the form factors $F_E$, $F_M$, and $F_Q$ and the
$A_i$ amplitudes is

\begin{equation}
 A_1 = F_E + {2 \over 3} F_Q
\end{equation}

\begin{equation}
 A_2 = (2(1 - \eta))^{-1} \lbrack ( 1 - {2 \over 3} \eta) F_Q
	 - F_E + F_M \rbrack
\end{equation}

\begin{equation}
 A_3 = F_M
\end{equation}

where

\begin{equation}
 \eta = { p^2 \over {4 m_{\rho}} }
\end{equation}

$p$ being $ (0,0,0,E_{cm}) $

\subsection{\bf Incorporating ${\rho}$ Decay in Amplitudes }

So far, the propagation and decay of the $\rho$'s has been
neglected.  This adds a Breit-Wigner factor for each $\rho$
and a factor $\epsilon_{\rho} \cdot Q_{\rho} $.
$Q_{\rho^{\pm}} = P_{\pi^{\pm}} - P_{\pi^0} $ where
the pions are the decay products
of the $\rho$ meson.  In addition,
the polarization of the intermediate $\rho$ must be summed over
using the completeness relation.

Thus, the matrix elements become

\begin{eqnarray}
 ME_1 = &\sum_{\lambda_{\rho^+} \lambda_{\rho^-} }
	  ( \epsilon_{\gamma^*} \cdot Q )
	  ( \epsilon_{\rho^+} \cdot \epsilon_{\rho^-} )
	  (\epsilon_{\rho^+} \cdot Q_{\rho^+})
	  (\epsilon_{\rho^-} \cdot Q_{\rho^+}) \nonumber \\
	  &{ 1 \over { (m^2_{\rho^+} - m^2_0 - i m_{\rho^+}
\Gamma_{\rho^+}) } }
	   { 1 \over { (m^2_{\rho^-} - m^2_0 - i m_{\rho^-}
\Gamma_{\rho^-}) } }
\end{eqnarray}

Now perform the summation over the polarizations of the
$ \rho^+ $ and $ \rho^- $.   Recall the relations

\begin{equation}
 \sum_{\lambda}
(  A \cdot \epsilon(\lambda) ) ( B \cdot
     \epsilon(\lambda)) =  - ( A \cdot  B)
     +  {( A \cdot  P) ( B \cdot  P)
	\over { M^2 } }
\end{equation}

\begin{eqnarray}
\sum_{\lambda_a \lambda_b} ( A \cdot \epsilon_a)
   (\epsilon_a \cdot \epsilon_b) (\epsilon_b \cdot  B)
   =
   &{ \Bigl[ -( { A} \cdot { B} ) + { { ({ A} \cdot P_{b})
( P_{b} \cdot { B}) } \over { P^2_{b} } } \Bigr] } \nonumber \\
   & - { ( A \cdot P_a) \over
   P_a^2 }({{ \Bigl[ -( {P_a} \cdot { B} ) + { { ({P_a}
\cdot P_{b}) ( P_{b} \cdot { B}) } \over { P^2_{b} } } \Bigr] }
})
\end{eqnarray}

Using these relations, the full matrix elements with the
$\rho$ decay become:

\begin{eqnarray}
 ME_1 = &(\epsilon_{\gamma^*} \cdot Q)
	 \Biggl\{
	  { \Bigl[ -( {Q_{\rho^+}} \cdot {Q_{\rho^-}} ) +
{ { ({Q_{\rho^+}} \cdot P_{\rho^-}) ( P_{\rho^-}
\cdot {Q_{\rho^-}}) } \over { P^2_{\rho^-} } } \Bigr] }
\nonumber      \\
	 & - { (Q_{\rho^+} \cdot P_{\rho^+} ) \over
	    P_{\rho^+}^2 }
	  ({ { \Bigl[ -( {P_{\rho^+}} \cdot {Q_{\rho^-}} ) +
{ { ({P_{\rho^+}} \cdot P_{\rho^-}) ( P_{\rho^-}
\cdot {Q_{\rho^-}}) } \over { P^2_{\rho^-} } } \Bigr] }
} ) \Biggr\} \nonumber \\
	  & { 1 \over { (m^2_{\rho^+} - m^2_0 - i m_{\rho^+}
\Gamma_{\rho^+}) } }
	 { 1 \over { (m^2_{\rho^-} - m^2_0 - i m_{\rho^-}
\Gamma_{\rho^-}) } }
\end{eqnarray}

\begin{eqnarray}
ME_2 = &( \epsilon_{\gamma^*} \cdot Q)
	( { \Bigl[ -( {Q} \cdot {Q_{\rho^+}} ) +
{ { ({Q} \cdot P_{\rho^+}) ( P_{\rho^+} \cdot {Q_{\rho^+}}) }
\over { P^2_{\rho^+} } } \Bigr] } )\nonumber \\
	&( { \Bigl[ -( {Q} \cdot {Q_{\rho^-}} ) +
{ { ({Q} \cdot P_{\rho^-}) ( P_{\rho^-} \cdot {Q_{\rho^-}}) }
\over { P^2_{\rho^-} } } \Bigr] } )\nonumber \\
	&  { 1 \over { (m^2_{\rho^+} - m^2_0 - i m_{\rho^+}
\Gamma_{\rho^+}) } }
	 { 1 \over { (m^2_{\rho^-} - m^2_0 - i m_{\rho^-}
\Gamma_{\rho^-}) } }
\end{eqnarray}

\begin{eqnarray}
ME_3 = &( { \Bigl[ -( {\epsilon_{\gamma^*}} \cdot {Q_{\rho^+}} )
+ { { ({\epsilon_{\gamma^*}} \cdot P_{\rho^+})
( P_{\rho^+} \cdot {Q_{\rho^+}}) }
\over { P^2_{\rho^+} } } \Bigr] } )\nonumber \\
	 & ( { \Bigl[ -( {Q} \cdot {Q_{\rho^-}} )
+ { { ({Q} \cdot P_{\rho^-}) ( P_{\rho^-} \cdot {Q_{\rho^-}}) }
\over { P^2_{\rho^-} } } \Bigr] } ) \nonumber \\
	 &+
	  ( { \Bigl[ -( {\epsilon_{\gamma^*}} \cdot {Q_{\rho^-}} )
+ { { ({\epsilon_{\gamma^*}} \cdot P_{\rho^-}) ( P_{\rho^-}
\cdot {Q_{\rho^-}}) } \over { P^2_{\rho^-} } } \Bigr] } )\nonumber \\
	 &( { \Bigl[ -( {Q} \cdot {Q_{\rho^+}} ) + { { ({Q}
\cdot P_{\rho^+}) ( P_{\rho^+} \cdot {Q_{\rho^+}}) }
\over { P^2_{\rho^+} } } \Bigr] } ) \nonumber \\
	& { 1 \over { (m^2_{\rho^+} - m^2_0 - i m_{\rho^+}
\Gamma_{\rho^+}) } }
	  { 1 \over { (m^2_{\rho^-} - m^2_0 - i m_{\rho^-}
\Gamma_{\rho^-}) } }
\end{eqnarray}

using these relations, the final expression for the amplitude
for the $\rho^+ \rho^-$ mode(s) becomes

\begin{eqnarray}
\Psi_{\rho^+ \rho^-} (\Omega)=
	  &  R_1 e^{i\phi_1} ( ME_1(\pi^0_1 \pi^0_2) +
	    ME_1(\pi^0_2 \pi^0_1) ) \nonumber \\
	 &+ R_2 e^{i\phi_2} ( ME_2(\pi^0_1 \pi^0_2) +
	    ME_2(\pi^0_2 \pi^0_1) )  \nonumber \\
	 &+ R_3 e^{i\phi_3} ( ME_3(\pi^0_1 \pi^0_2) +
	    ME_3(\pi^0_2 \pi^0_1) )
\end{eqnarray}

$ME_i(\pi^0_1 \pi^0_2)$ and $ ME_i(\pi^0_2 \pi^0_1)$ refer to the two
choices of $\pi^0$ to form the 4-vectors in order to satisfy Bose
statistics under interchange of the two identical neutral pions.
Defining $\pi^0_1$ as the more energetic neutral pion (e.g. $
E_{\pi^0_1} \ge E_{\pi^0_2} $) in the center of mass of the four
pions frame of reference provides a well-defined and unique way
to label the two neutral pions in real data.

\subsection{\bf Simplification of $ \rho^+ \rho^- $ Amplitudes }

The preceding amplitudes are complicated.
However, only a few terms dominate the amplitudes.
The factors $ ( Q_{\rho^{\pm}}
\cdot P_{\rho^{\pm}}) $ in the formulas above are the reason for
this.

\begin{equation}
 (Q_{\rho^{\pm}}  \cdot P_{\rho^{\pm} } ) =
     (P_{\pi^{\pm}} - P_{\pi^0}) \cdot
     (P_{\pi^{\pm}} + P_{\pi^0})
\end{equation}

This becomes

\begin{equation}
 (Q_{\rho^{\pm}}  \cdot P_{\rho^{\pm} } ) =
      ( P_{\pi^{\pm} } \cdot P_{\pi^{\pm} } ) -
      ( P_{\pi^0 } \cdot P_{\pi^0 } )
\end{equation}

which becomes

\begin{equation}
	 (Q_{\rho^{\pm}}  \cdot P_{\rho^{\pm} } ) =
	 m_{\pi^{\pm}}^2 - m_{\pi^0}^2
\end{equation}

If the masses of the charged and neutral pions were the
same, this term would be exactly zero.  In this case the
$\rho^+ \rho^-$ amplitudes reduce to

\begin{eqnarray}
ME_1 = &- (\epsilon_{\gamma^*} \cdot Q)
	    (Q_{\rho^+} \cdot Q_{\rho^-} ) \nonumber \\
	   & { 1 \over { (m^2_{\rho^+} - m^2_0
- i m_{\rho^+} \Gamma_{\rho^+}) } }
	     { 1 \over { (m^2_{\rho^-} - m^2_0
- i m_{\rho^-} \Gamma_{\rho^-}) } }
\end{eqnarray}

\begin{eqnarray}
ME_2 = &( \epsilon_{\gamma^*} \cdot Q) \Biggl(
	  -\Bigl( (Q \cdot Q_{\rho^+}) +
		 (Q \cdot Q_{\rho^-}) \Bigr) \Biggr) \nonumber \\
	    & { 1 \over { (m^2_{\rho^+} - m^2_0
- i m_{\rho^+} \Gamma_{\rho^+}) } }
	      { 1 \over { (m^2_{\rho^-} - m^2_0
- i m_{\rho^-} \Gamma_{\rho^-}) } }
\end{eqnarray}

\begin{eqnarray}
ME_3 = & - \Bigl(
( \epsilon_{\gamma^*} \cdot Q_{\rho^+} )
	  (Q \cdot Q_{\rho^-} ) +
	  ( \epsilon_{\gamma^*} \cdot Q_{\rho^-} )
	  (Q \cdot Q_{\rho^+} ) \Bigr)  \nonumber \\
	 &   { 1 \over { (m^2_{\rho^+} - m^2_0
- i m_{\rho^+} \Gamma_{\rho^+}) } }
	     { 1 \over { (m^2_{\rho^-} - m^2_0
- i m_{\rho^-} \Gamma_{\rho^-}) } }
\end{eqnarray}

Although the masses of the charged and neutral
pions are slightly different, the terms above
dominate the $\rho^+ \rho^-$ amplitudes.  Another
consequence is that the $\rho^+ \rho^-$ amplitudes are
largely insensitive to the choice of $1 \over M^2 $ or
$ 1 \over P^2 $ in the completeness relation.

\subsection{\bf The $e^+ e^- \rightarrow a_1 \pi$ Mode }

This process consists of
$$ e^+ e^- \rightarrow a_1^{\pm} \pi^{\mp} $$
where
$$ a_1^{\pm} \rightarrow \rho^{\pm} \pi^0 $$
and
$$\rho^{\pm} \rightarrow \pi^{\pm} \pi^0 $$

At the $e^+ e^- \rightarrow \gamma^* \rightarrow a_1 \pi $
step the relevant 4-vectors are $\epsilon_{\gamma}$,
$ \epsilon_{a_1}$, $P = P_{a_1} + P_{\pi} $ and
$Q = P_{a_1} - P_{\pi} $.  There are two independent
amplitudes:

\begin{equation}
ME_{e^+ e^- \rightarrow a_1 \pi}^I = ( {\epsilon_{\gamma^*}}
\cdot \epsilon_{a_1} )
\end{equation}

\begin{equation}
ME_{e^+ e^- \rightarrow a_1 \pi}^{II} = ({\epsilon_{\gamma^*}}
\cdot Q) (\epsilon_{a_1} \cdot P)
\end{equation}

At the $a_1$ vertex $ a_1 \rightarrow \rho \pi$ there are
four relevant 4-vectors: $\epsilon_{a_1}$, ${\epsilon_{\rho}}$,
$ P_{a_1} $, and $ Q_{a_1} = P_{\rho} - P_{\pi} $.  These
are formed into two independent amplitudes:

\begin{equation}
ME_{a_1 \rightarrow \rho \pi}^I = ( {\epsilon_{a_1} } \cdot
{\epsilon_{\rho}} )
\end{equation}

\begin{equation}
ME_{a_1 \rightarrow \rho \pi}^{II} = ( {\epsilon_{a_1} } \cdot
Q_{a_1} ) ( {\epsilon_{\rho}}
\cdot  P_{a_1})
\end{equation}

At the $\rho$ vertex, there is only one matrix element

\begin{equation}
 ME_{\rho \rightarrow \pi \pi} = ({\epsilon_{\rho}} \cdot  Q_{\rho} )
\end{equation}

as seen before.

Combining all of these yields four independent amplitudes for
the $a_1 \pi $ process.  These are

\begin{equation}
 ME^I_{a_1 \pi} = \sum_{\lambda_{a_1} \lambda_{\rho}}
({\epsilon_{\gamma^*}} \cdot
{\epsilon_{a_1} }) ({\epsilon_{a_1} } \cdot {\epsilon_{\rho}})
({\epsilon_{\rho}} \cdot  Q_{\rho})
\end{equation}

\begin{equation}
 ME^{II}_{a_1 \pi} = \sum_{\lambda_{a_1} \lambda_{\rho}}
({\epsilon_{\gamma^*}} \cdot
{\epsilon_{a_1} }) ({\epsilon_{a_1} } \cdot  Q_{a1})
({\epsilon_{\rho}} \cdot P_{a_1}) ({\epsilon_{\rho}} \cdot
Q_{\rho})
\end{equation}

\begin{equation}
 ME^{III}_{a_1 \pi} = \sum_{\lambda_{a_1} \lambda_{\rho}}
({\epsilon_{\gamma^*}} \cdot
 Q) ({\epsilon_{a_1} } \cdot  P ) ({\epsilon_{a_1} } \cdot
{\epsilon_{\rho}}) ({\epsilon_{\rho}} \cdot  Q_{\rho})
\end{equation}

\begin{equation}
 ME^{IV}_{a_1 \pi} = \sum_{\lambda_{a_1} \lambda_{\rho}}
({\epsilon_{\gamma^*}} \cdot
 Q) ( {\epsilon_{a_1} } \cdot  P ) ({\epsilon_{a_1} }
\cdot  Q_{a_1}) ({\epsilon_{\rho}} \cdot  P_{a_1})
({\epsilon_{\rho}} \cdot  Q_{\rho} )
\end{equation}

Now, performing the sums over the polarizations yields:

\begin{equation}
	ME^{I}_{a_1 \pi} = \sum_{\lambda_{\rho}}
{ \Bigl[ -( {\epsilon_{\gamma^*}} \cdot {\epsilon_{\rho}} )
+ { { ({\epsilon_{\gamma^*}} \cdot P_{a_1})
( P_{a_1} \cdot {\epsilon_{\rho}})
 } \over { P^2_{a_1} } } \Bigr] }
	({\epsilon_{\rho}} \cdot  Q_{\rho})
\end{equation}

\begin{equation}
	ME^{II}_{a_1 \pi} =
{ \Bigl[ -( {{\epsilon_{\gamma^*}} } \cdot {
Q_{a_1}} ) + { { ({{\epsilon_{\gamma^*}} } \cdot P_{a_1}) ( P_{a_1}
\cdot { Q_{a_1}}) } \over { P^2_{a_1} } } \Bigr] }
	  { \Bigl[ -( { P_{a_1}} \cdot { Q_{\rho}} ) + { { ({ P_{a_1}
} \cdot P_{\rho}) ( P_{\rho} \cdot { Q_{\rho}}) } \over { P^2_{\rho}
} } \Bigr] }
\end{equation}

\begin{equation}
	ME^{III}_{a_1 \pi} =
\sum_{\lambda_{\rho}} ({\epsilon_{\gamma^*}}
  \cdot  Q)
		{ \Bigl[ -( { P} \cdot {{\epsilon_{\rho}}} ) + { { ({
 P} \cdot P_{a_1}) ( P_{a_1} \cdot {{\epsilon_{\rho}}}) } \over { P^2
_{a_1} } } \Bigr] }
	  ({\epsilon_{\rho}} \cdot  Q_{\rho} )
\end{equation}

\begin{eqnarray}
\label{eq:a_one_me_iv}
	ME^{IV}_{a_1 \pi} = & ({\epsilon_{\gamma^*}}  \cdot  Q)
       { \Bigl[ -( { P} \cdot { Q_{a_1}} ) + { { ({ P} \cdot P_{a_1})
 ( P_{a_1} \cdot { Q_{a_1}}) } \over { P^2_{a_1} } } \Bigr] }
\nonumber \\
       & { \Bigl[ -( { P_{a_1}} \cdot { Q_{\rho}} )
+ { { ({ P_{a_1}} \cdot P_{\rho}) ( P_{\rho} \cdot { Q_{\rho}}) }
\over { P^2_{\rho} } }
 \Bigr] }
\end{eqnarray}

Performing the sums over the polarizations of the $\rho$ deferred
in two of the four yields:

\begin{eqnarray}
\label{eq:a_one_me_one}
ME^{I}_{a_1 \pi} =
& - \Bigl[ - ( {{\epsilon_{\gamma^*}} } \cdot { Q_{\rho
}} ) + { {( {{\epsilon_{\gamma^*}} } \cdot P_{\rho} )
( P_{\rho} \cdot { Q_{\rho}} ) } \over { P^2_{\rho}} } \Bigr]
\nonumber \\
& + { { \epsilon_{\gamma
^*} \cdot { P_{a_1}}} \over { P^2_{a_1}} }{ \Bigl[ -( {{P_{a_1}} }
\cdot { Q_{\rho}}) + { {({{P_{a_1}} } \cdot P_{\rho})(P_{\rho}
\cdot {
Q_{\rho}}) } \over {P^2_{\rho} } } \Bigr]}
\end{eqnarray}

\begin{eqnarray}
ME^{II}_{a_1 \pi} =
{ \Bigl[ -( {{\epsilon_{\gamma^*}} } \cdot { Q_{a_1}}
 ) + { { ({{\epsilon_{\gamma^*}} } \cdot P_{a_1}) ( P_{a_1}
\cdot { Q_{a_1}}) } \over { P^2_{a_1} } } \Bigr] }
	  { \Bigl[ -( { P_{a_1}} \cdot { Q_{\rho}} ) + { { ({ P_{a_1}
} \cdot P_{\rho}) ( P_{\rho} \cdot { Q_{\rho}}) } \over { P^2_{\rho}
} } \Bigr] }
\end{eqnarray}

\begin{eqnarray}
	   ME^{III}_{a_1 \pi} = & ({\epsilon_{\gamma^*}}  \cdot  Q)
	  \Biggl[ - \Bigl[ - ( { P} \cdot { Q_{\rho}} ) + { {( { P}
\cdot P_{\rho} ) ( P_{\rho} \cdot { Q_{\rho}} ) } \over { P^2_{\rho}}
 } \Bigr] \nonumber \\
& + { { P \cdot { P_{a_1}}} \over { P^2_{a_1}} }%
{ \Bigl[ -( {P_{a_1}} \cdot { Q_{\rho}}) + %
{ {({P_{a_1}} \cdot P_{\rho})(P_{\rho} \cdot { Q_{\rho}}) }%
 \over {P^2_{\rho} } } \Bigr]}  \Biggr]
\end{eqnarray}

\begin{eqnarray}
	ME^{IV}_{a_1 \pi} = ({\epsilon_{\gamma^*}}  \cdot  Q)
       { \Bigl[ -( { P} \cdot { Q_{a_1}} ) + { { ({ P} \cdot P_{a_1})
 ( P_{a_1} \cdot { Q_{a_1}}) } \over { P^2_{a_1} } } \Bigr] }
       { \Bigl[ -( { P_{a_1}} \cdot { Q_{\rho}} ) + { { ({ P_{a_1}}
 \cdot P_{\rho}) ( P_{\rho} \cdot { Q_{\rho}}) }
\over { P^2_{\rho} } }
 \Bigr] }
\end{eqnarray}

These matrix elements must be multiplied by Breit-Wigner formulas for
$a_1$ and $\rho$ and Bose symmetrized.

\begin{eqnarray}
\label{eq:a_one_all}
\Psi_{a_1 \pi}(\Omega) =
& R_1 e^{i\phi_1} ( ME^I_{a_1}(\pi^0_1 \pi^0_2)
 + ME^I_{a_1 \pi} (\pi^0_2 \pi^0_1) ) \nonumber \\
& + R_2 e^{\phi_2} ( ME^{II}_{a_1}(\pi^0_1 \pi^0_2)
 + ME^{II}_{a_1 \pi} (\pi^0_2 \pi^0_1) ) \nonumber \\
& + R_3 e^{\phi_3} ( ME^{III}_{a_1}(\pi^0_1 \pi^0_2)
 + ME^{III}_{a_1 \pi} (\pi^0_2 \pi^0_1) ) \nonumber \\
& + R_4 e^{\phi_4} ( ME^{IV}_{a_1}(\pi^0_1 \pi^0_2)
 + ME^{IV}_{a_1 \pi} (\pi^0_2 \pi^0_1) )
\end{eqnarray}

\subsection{\bf The $e^+ e^- \rightarrow \omega^0 \pi^0 $ Mode }

This formalism can also provide a formula for the process
$$ e^+ e^- \rightarrow \omega^0 \pi^0 $$
$$ \omega^0 \rightarrow \pi^+ \pi^- \pi^0 $$

The $ e^+ e^- \rightarrow \omega \pi $ vertex has
the matrix element:

\begin{equation}
 ME_{\omega \pi} = \epsilon_{\mu \nu \rho \sigma}
\epsilon^{\mu}_{\gamma^*}
		 \epsilon_{\omega}^{\nu}
		 P^{\rho}
		 Q^{\sigma}
\end{equation}

where $P = P_{\omega} + P_{\pi^0} $ and $ Q = P_{\omega} -
P_{\pi^0} $

The $ \omega \rightarrow \pi^+ \pi^- \pi^0 $ vertex is
governed by the matrix element.

\begin{equation}
 ME_{\pi \pi \pi} = {\epsilon_{\mu \nu \rho \sigma }}
{\epsilon_{\omega}^{\mu}} P_{\pi^+}^{\nu}
	 P_{\pi^-}^{\rho}  P_{\omega}^{\sigma}
\end{equation}

Since $P_{\omega} = P_{\pi^+} + P_{\pi^-}
				 + P_{\pi^0} $, this formula
is equivalent to

\begin{equation}
ME_{\pi \pi \pi} = {\epsilon_{\mu \nu \rho \sigma }}
{\epsilon_{\omega}^{\mu}} P_{\pi^+}^{\nu}
	 P_{\pi^-}^{\rho}  P_{\pi^0}^{\sigma}
\end{equation}

The complete matrix element is

\begin{eqnarray}
ME_{\omega \pi} = & \sum_{\lambda_{\omega} }
		     {\epsilon_{\mu \nu \rho \sigma }}
		     {\epsilon_{\gamma^*}} ^{\mu}
		     \epsilon_{\omega}^{\nu}
		     P^{\rho}
		      Q^{\sigma} \\
		&  { \epsilon_{\alpha \beta \gamma \delta}}
		    \epsilon_{\omega}^{\alpha}
	    P_{\pi^+}^{\beta} P_{\pi^-}^{\gamma} P_{\omega}^{\delta}
		    { 1 \over { (m^2_{\omega} - m^2_0
- i m_{\omega} \Gamma_{\omega}) } }
\end{eqnarray}

This formula reduces to a simple formula in terms of dot products of
4-vectors (polarization 4-vectors and 4-momenta).

First, use the completeness relation for the $\omega$ to
get

\begin{eqnarray}
ME_{\omega \pi} = &\epsilon^{\mu \nu \rho \sigma}
		     P_{\mu} Q_{\nu} \epsilon^{\gamma^*}_{\rho}
		     \epsilon^{\alpha \beta \gamma \delta}
		     P^{\omega}_{\beta} P^{\pi^+}_{\gamma}
		     P^{\pi^-}_{\delta} \nonumber \\
		   &  ( - g_{\sigma \alpha} + { P^{\omega}_{\sigma}
			 P^{\omega}_{\alpha} \over M^2_{\omega} } )
{ 1 \over
{ (m^2_{\omega} - m^2_0 - i m_{\omega} \Gamma_{\omega}) } }
\end{eqnarray}

Note that:

\begin{equation}
{\epsilon^{\alpha \beta \gamma \delta} P^{\omega}_{\beta}
   P^{\pi^+}_{\gamma} P^{\pi^-}_{\delta} P^{\omega}_{\sigma}
   P^{\omega}_{\rho} \over { M^2_{\omega} } } = 0
\end{equation}

since $P^{\omega}$ is repeated.  Thus the matrix element becomes
the simpler form:

\begin{eqnarray}
ME_{\omega \pi} = &\epsilon^{\mu \nu \rho \sigma}
		     P_{\mu} Q_{\nu} \epsilon^{\gamma^*}_{\rho}
		     \epsilon^{\alpha \beta \gamma \delta}
		     P^{\omega}_{\beta} P^{\pi^+}_{\gamma}
		     P^{\pi^-}_{\delta} \nonumber \\
		  &  ( - g_{\sigma \alpha} )
{ 1 \over { (m^2_{\omega} - m^2_0
- i m_{\omega} \Gamma_{\omega}) } }
\end{eqnarray}

At this point, use the relation

\begin{equation}
\epsilon^{\mu \nu \rho \sigma } \epsilon_{\rho \beta \gamma \delta}
     = \epsilon^{\rho \mu \nu \sigma} \epsilon_{\rho \beta \gamma
       \delta} = - \delta^{\mu \nu \sigma}_{\beta \gamma \delta}
\end{equation}

where

\begin{equation}
\delta^{abc}_{def} = \left\{  \begin{array}{cc}
		0 & \mbox{if index repeated in $abc$ or $def$} \\
		1 & \mbox{if $def$ is even permutation of $abc$} \\
		-1 & \mbox{if $def$ is odd permutation of $abc$}
			\end{array}
		\right.
\end{equation}

This is the Kronecker 3-delta.  It is a generalization of
the Kronecker delta $ \delta^i_j $.

\begin{equation}
 \delta^a_b = \left\{  \begin{array}{cc}
				1  & \mbox{if $a = b$} \\
				0  & \mbox{otherwise}
			\end{array}
		\right.
\end{equation}

The Kronecker 2-delta is defined by

\begin{equation}
	\delta^{ab}_{cd} = \left\{ \begin{array}{cc}
		1 & \mbox{if $ab$ is even permutation of $cd$} \\
		-1 & \mbox{if $ab$ is odd permutation of $cd$} \\
			0 & \mbox{if index repeated in $ab$ or $cd$}
			\end{array}
		\right.
\end{equation}

The Kronecker 2-delta can be expressed in terms of the Kronecker
delta:

\begin{equation}
 \delta^{ab}_{cd} = ( \delta^a_c \delta^b_d -
			\delta^a_d \delta^b_c )
\end{equation}

The Kronecker 3-delta can be expanded in terms of the simple
Kronecker delta as well:

\begin{equation}
\delta^{abc}_{def} =  ( \delta^a_d \delta^b_e \delta^c_f
			- \delta^a_e \delta^b_d \delta^c_f
			- \delta^a_d \delta^b_f \delta^c_e
			- \delta^a_f \delta^b_e \delta^c_d
			+ \delta^a_e \delta^b_f \delta^c_d
			+ \delta^a_f \delta^b_d \delta^c_e )
\end{equation}

This leads to an expansion in terms of dot products:

\begin{eqnarray}
A_{\mu} B_{\nu} C_{\rho}
\delta^{\alpha \beta \gamma}_{\mu \nu \rho} D^{\alpha}
E^{\beta} F^{\gamma}  = & ( A \cdot  D) (  B \cdot  E )
   ( C \cdot  F)
   - ( B \cdot  D) ( E \cdot  A)
     ( C \cdot  F) \nonumber \\
   & - ( A \cdot  D) ( C \cdot  E)
     ( B \cdot  F)
   - ( C \cdot  D) ( B \cdot  E)
     ( A \cdot  F) \nonumber \\
   & + ( B \cdot  D) ( C \cdot  E)
     ( A \cdot  F)
   + ( C \cdot  D) ( A \cdot  E)
     ( B \cdot  F)
\end{eqnarray}

This can be used to produce a form for the $\omega \pi$
matrix element in terms of dot products of the
4-momenta and the virtual photon polarization 4-vector.

\begin{eqnarray}
ME = &{\epsilon_{\gamma^*}}  \cdot \lbrack
	    ( P \cdot  P_{\pi^-})
	    ( Q \cdot  P_{\omega})  P_{\pi^+}\nonumber \\
	& -  ( P \cdot  P_{\omega})
	    ( Q \cdot  P_{\pi^-})  P_{\pi^+} \nonumber \\
	& -  ( P \cdot  P_{\pi^+})
	    ( Q \cdot  P_{\omega})  P_{\pi^-} \nonumber \\
	& +  ( P_{\omega} \cdot  P)
	    ( Q \cdot  P_{\pi^+})  P_{\pi^-}  \nonumber \\
	& +  ( P \cdot  P_{\pi^+})
	    ( Q \cdot  P_{\pi^-})  P_{\omega} \nonumber \\
	& -  ( P \cdot  P_{\pi^-})
	    ( Q \cdot  P_{\pi^+})  P_{\omega}
	  \rbrack \nonumber \\
& { 1 \over
{ (m^2_{\omega} - m^2_0 - i m_{\omega} \Gamma_{\omega}) } }
\end{eqnarray}

\section{Conclusion}

Some general formulas for simplifying Lorentz invariant amplitude
expressions were developed.  These were applied to calculate the
Lorentz invariant amplitudes for processes contributing to
\({e^+ e^- \rightarrow \pi^+ \pi^- \pi^0 \pi^0}\).
These formula are also applicable to
\(e^+ e^- \rightarrow \pi^+ \pi^- \pi^+ \pi^-\),
\(e^+ e^- \rightarrow
K^+ K^- \pi^0 \pi^0\), \(e^+ e^- \rightarrow K^+ K^- K^0 K^0 \), and
\(e^+ e^- \rightarrow K^+ K^- K^+ K^- \).
These formulas may be used to determine
the electromagnetic form factors of the \(\rho\) meson and
properties of
the \(a_1\) resonance.

\section{Acknowledgements}

I wish to acknowledge the assistance of the other members of the
Illinois group at the Stanford Linear Accelerator Center (SLAC): my
advisor, Jon J. Thaler, Gary Gladding, Bob Eisenstein, Joe Izen,
Basil
Tripsas, Tim Freese, and Walid Majid.  I also wish to acknowledge a
number of people with whom I worked on the SLD project at SLAC: Ron
Cassell,
Sridhara Dasu, Nety Krishna, Gary Bower, Ray Cowan, and Terry Reeves.
I want to thank Bill Lockman for his considerable
advice regarding the
analysis in the thesis on which this paper is based.  Several others
provided valuable advice and discussion of the analysis: Bill
Dunwoodie, Liang Peng Chen, Walter Toki, Larry Parrish, and Jon Labs.
Ramon Berger provided invaluable assistance by
keeping the {\bf Mark III}\
software system operational on the IBM mainframes at SLAC.

This research was supported in part
by the U.S. Department of Energy, under
contract DE-AC02-76ER01195 and contract DE-AC02-76ER40677.


\begin{table}[tbh]
\caption{Isobar Model Modes}
\label{tab:modes}
\begin{tabular}{c}
Mode  \\  \hline
\( e^+ e^- \rightarrow \omega \pi \) \\
\( e^+ e^- \rightarrow \rho^+ \rho^- \) \\
\( e^+ e^- \rightarrow a_1 \pi \) \\
\( e^+ e^- \rightarrow \pi^+ \pi^- \pi^0 \pi^0 \) \\ 
\end{tabular}
\end{table}


\begin{references}

\bibitem{ft:Watson-1}{K.M.~Watson, {\it Phys. Rev.} {\bf 88}, 1163
(1952) }

\bibitem{ft:Watson-2}{K.M.~Watson, {\it Phys. Rev.} {\bf 95}, 228
(1954) }

\bibitem{ft:Cook} {L.F.~Cook and B.W.~Lee, {\it Phys. Rev.}
{\bf 127}, 283
 (1962) }

\bibitem{ft:Ball}{J.S.~Ball, W.R.~Frazer, and M.~Nauenberg,
{\it Phys. Rev.}
 {\bf 128}, 478 (1962) }

\bibitem{ft:Mandelstam}{S.~Mandelstam, J.E.~Paton, R.F.~Peierls, and
 A.Q.~Sarker, {\it Ann. Phys. } {\bf 18}, 198 (1962) }

\bibitem{ft:Fleming}{G.N.~Fleming, {\it Phys. Rev.} {\bf B135}, 551
(1964) }

\bibitem{ft:Freedman}{D.Z.~Freedman, C.~Lovelace,
and J.M.~Namyslowski,
{\it Nuovo Cimento} {\bf 43A }, 258 (1966) }

\bibitem{ft:Morgan}{D.~Morgan, {\it Phys. Rev. } {\bf 166}, 1731
(1968) }

%
\bibitem{ft:Webster}{V.~Neufeldt and D.~B.~Guralnik, { \it Webster's
New World Dictionary of American
English }, ( Simon and Schuster, New York, 1986 ) p. 1013}

\bibitem{ft:LePage} G.~Peter LePage and
Stanley J. Brodsky, {\it Phys. Rev.}
 {\bf D22}, 2157 (1980)

\bibitem{ft:Brodsky} { Stanley J. Brodsky
and G.~Peter LePage, {\it Phys. Rev.}
{\bf D24 }, 2848 (1981) }

%

\bibitem{ft:Chernyak-1}{ V.L.~Chernyak
and A.R.~Zhitnitsky, {\it Nuc. Phys. }
{\bf B201}, 492 (1982) }
\bibitem{ft:Chernyak-2}{ V.L.~Chernyak, A.R.~Zhitnitsky,
and I.R.~Zhitnitsky,
 {\it Nuc. Phys. } {\bf B204}, 477 (1982) }

%
%
\bibitem{ft:McGowan}{ J.F.~McGowan, Ph.~D. Thesis,
University of Illinois at Urbana-Champaign, UMI-93-14914-mc}
%
%
%

\bibitem{ft:Wheeler}{J.A.~Wheeler, {\it Phys. Rev. }
{\bf 52 }, 1107 (1937) }

\bibitem{ft:Heisenberg}{W.~Heisenberg, {\it Z. Physik }
{\bf 120}, 513, 673
 (1943) }

%
%
\bibitem{ft:Blatt} { J.~Blatt and V.~Weisskopf,
{\it Theoretical Nuclear
 Physics}, (John Wiley and Sons, New York, 1952) p. 332}
%
%

\bibitem{ft:Kramer}   G.~Kramer and T.F.~Walsh,
{\it Quasi Two Body $ e^+ e^- $  Annihilation },{\it Z. Physik }
{\bf 263}, 361-386 (1973)
\end{references}
\end{document}